# Discontinuous transitions: Multiphasic profiles for channels, binding, pH, folding and chain length

Per Nissen

Norwegian University of Life Sciences

Department of Ecology and Natural Resource Management

P. O. Box 5003, NO-1432 Ås, Norway

per.nissen@nmbu.no

2015

# Abstract

A wide variety of biological as well as non-biological processes and phenomena involving ion channels, binding, pH, folding/unfolding and effects of chain length are well represented by multiphasic profiles, a series of straight lines separated by discontinuous transitions, often in the form of noncontiguities (jumps). The lines appear to be perfectly straight, with no indication of any curvilinearity. Adjacent lines are quite often parallel or nearly so and are then necessarily separated by jumps. If the data are sufficiently detailed and precise, the conventional curvilinear profiles must be rejected for a variety of reasons, whereas sudden changes of state occur in many physical systems (see Conclusions).

# Introduction

Multiphasic profiles, i.e. profiles representing discrete states, were first found for the concentration dependence, plotted in linear transformations of the Michaelis-Menten equation, for uptake of sulfate (Nissen 1971) and other ions in plants (see Nissen 1974, 1991, 1996) for reviews. It turns out than not only conductance-voltage and conductance-pressure data for ion channels, but also data for many other processes (binding as determined by isothermal calorimetry or anisotropy, pH dependences, and chemical and thermal folding/unfolding and scattering of proteins) can be well represented by multiphasic profiles. Plots of $\Delta G$ vs. $\Delta H$ or log $K_m$ vs log $k_{cat}$ for substrates with increasing chain length also give profiles with sharp transitions and marked jumps.

One or a few data sets for each of these processes are examined in this paper (using published tabular data or original data kindly provided by the authors) to see whether the profiles should be represented as multiphasic, or whether the conventional curvilinear interpretation remains valid. In addition to the r values, slopes $\pm$ SE (or only slopes) are given on the plots. As will be evident from the plots, the slopes are either given as such or have been multiplied by 1000. The Runs test, or Wald-Wolfowitz test (Wald and Wolfowitz 1940), a non-parametric test, gives the probability for the uneven distribution of points around the curvilinear profile being due to chance. Fisher's method for combining independent probabilities (Fisher 1954) has been used to determine the probability for two or more profiles.

# Reanalyses

## Ion channels

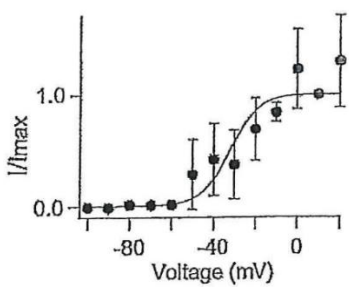

**Fig. 1.** Fig. S5d in Devaraneni et al. (2013). Voltage-dependent gating of the $K_vAP^{G198dA}$ channel. Tail currents were recorded by stepping to -100 mV following the test pulse. The fraction of the maximal current observed was plotted as a function of the test potential and fitted to the Boltzmann equation to give values of $V_{0.5} = 31.9 \pm 15$ mV and $z = 3.85 \pm 1.6$ (mean $\pm$ SD, $n = 3$). The smooth line corresponds to a Boltzmann function plotted with the average values. Original data kindly provided by Kimberly Matulef and Francis Valiyaveetil.

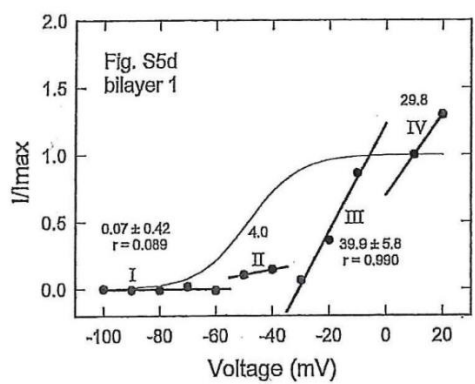

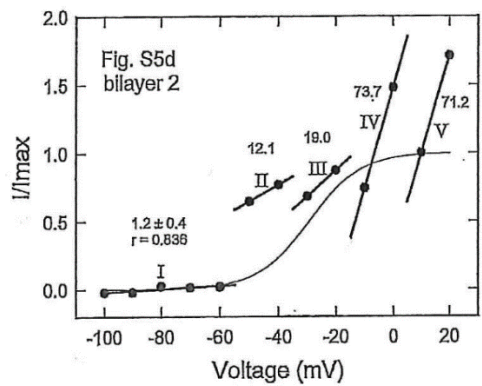

**Fig. 2** (left above). The data for bilayer 1 in Fig. 1 are well represented as tetraphasic, with the transitions between -60 and -50, between -40 and -30, and between -10 and 10 mV (the point for 0 mV is missing). All transitions are in the form of jumps. The points are very unevenly distributed around the curve (points 1-10 are below, points 11-12 are above, P = 3.0% by the Runs test).

**Fig. 3** (right above). The data for bilayer 2 are pentaphasic, with the transitions between -60 and -50, between -40 and -30, between -20 and -10, and between 0 and 10 mV. All transitions are in the form of jumps. Lines IV and V are parallel.

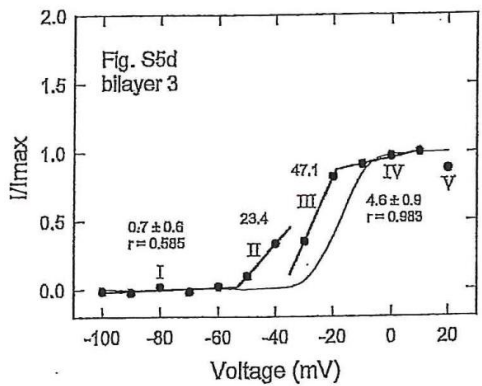

**Fig. 4**. The data for bilayer 3 can be well represented as pentaphasic, with the transitions at -53.4, between -40 and -30 (jump), at -18.9, and between 10 and 20 mV.

The multiphasic profiles for bilayers 1-3 differ quite markedly. The multiphasic profile for the average values will therefore be meaningless and is not shown.

The fits to the Boltzmann function are for the most part very poor for the data in Fig. 1. In contrast, multiphasic profiles give good fits. The many jumps clearly show that the profiles are discontinuous.

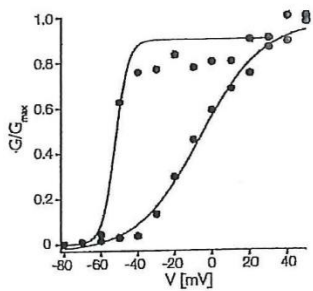

**Fig. 5.** Fig. S1 in Schmidt et al. (2012). Kv2.1 in outside-out patch (left) or whole-cell configuration (right) recorded from the same CHO cell. Boltzmann functions (solid lines) fit data with $V_m$ (mV) and z ($q_e$): -52.3 ± 1.42, 7.52 ± 2.7 (outside-out patch), and -6.32 ± 2, 1.5 ± 0.1 (whole cell). Original data kindlyprovided by Daniel Schmidt.

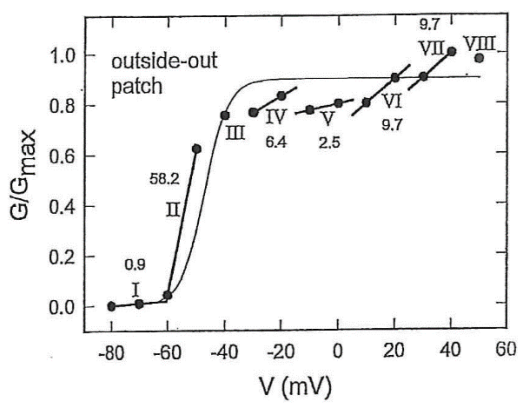

**Fig. 6.** The profile for the outside-out patch in Fig. 5 can be well represented by 8 phases, with the transitions at -60.4, on both sides of -40, between -20 and -30 (jump), between 0 and 10 (jump), between 20 and 30 (jump), and between 40 and 50 mV. The data are insufficiently detailed in the range of phases III and VIII for resolution of the lines. Lines VI and VII are parallel. Runs test: 2.8%.

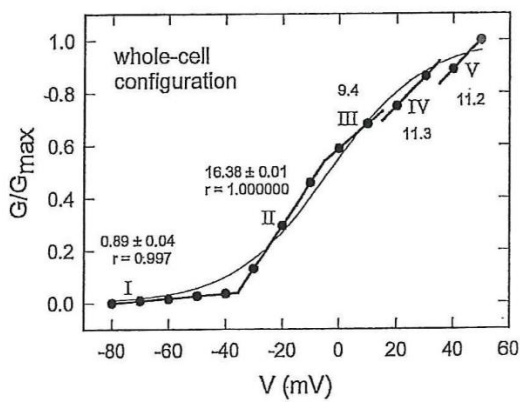

**Fig. 7.** The profile for the whole-cell configuration can be well represented as pentaphasic, with the transitions at -35.6 and -5.0, between 10 and 20 (jump), and between 30 and 40 mV (jump). The r value for line II is exceedingly high. Lines IV and V are parallel. Runs test: 6.8%.

The combined probability that the uneven distribution of points around the two curvilinear profiles in Fig. 5 is due to chance is less than 2.5%. The fits of the Boltzmann curves are also poor. In contrast, the multiphasic profiles give good fits, have parallel lines and lines with high to exceedingly high r values.

A reanalysis of the other current-voltage data (Figs lD, lH, lL, 2C, 4C, 4F, S2 and S4) in this paper will be given in another arXiv paper.

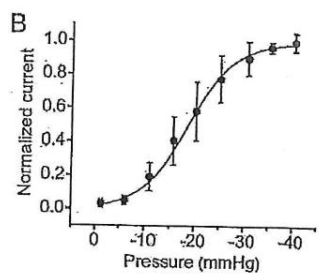

**Fig. 8.** Fig. S1B in Pathak et al. (2014). Current-pressure relationship of stretch-activated currents in an SC23 human neural stem/progenitor cell at a holding potential of -80 mV fitted with a Boltzmann equation with $P_{50}$ = -18.6 mm Hg, s = 4.7 mm Hg ($n$ = 5 cells, mean ± SE). Original data kindly provided by Medha M. Pathak and Francesco Tombola.

The curvilinear and multiphasic profiles for the 5 replicate experiments in Fig. 8 are shown in the following. (The x axis has not been inverted.)

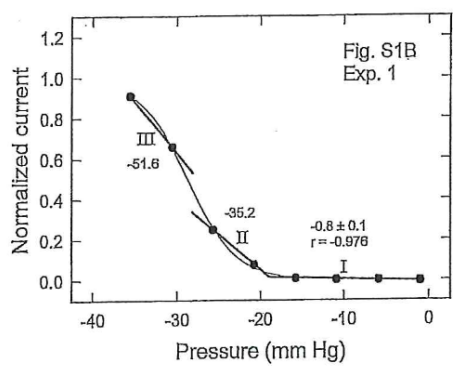

**Fig. 9.** Triphasic. Transitions at -18.8 and between -25.7 and -30.6 mm Hg (jump).

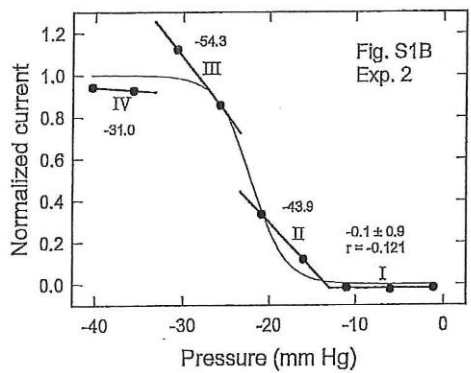

**Fig. 10.** Tetraphasic. Transitions at -12.9, between -20.9 and -25.8 (jump), and between -30.7 and -35.6 mm Hg (jump).

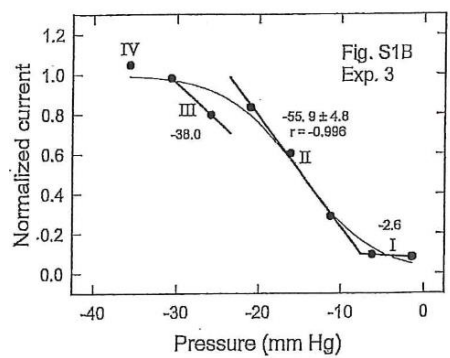

**Fig. 11.** Tetraphasic. Transitions at -7.6, between -21.0 and -25.9 (jump), and between -30.7 and -35.5 mm Hg.

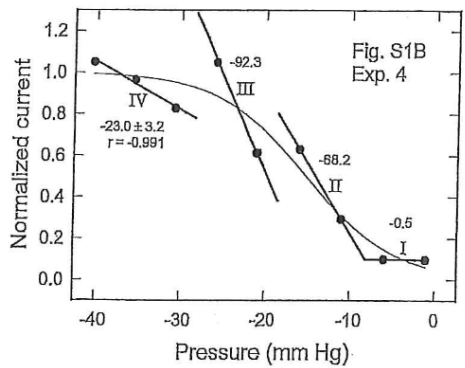

**Fig. 12.** Tetraphasic. Transitions at -8.2, between -16.0 and -20.9 (jump), and between -25.8 and -30.7 mm Hg (jump).

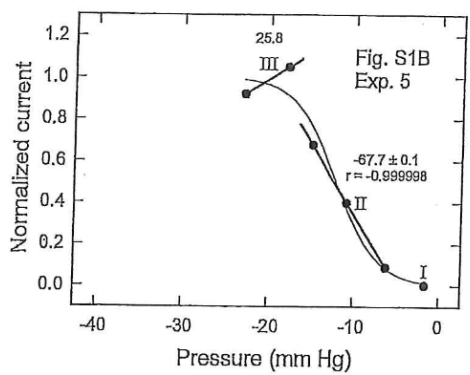

**Fig. 13.** Triphasic. Transitions between -1.6 and -6.2, and between -14.9 and -17.8 mm Hg (jump). Exceedingly high absolute r value for line II. Line III has a positive slope.

The fits to the Boltzmann curve are good for the data in experiment 1, but are in part very poor for the other four experiments. The one or two large jumps in the multiphasic profiles clearly show that the profiles are discontinuous.

**Binding**

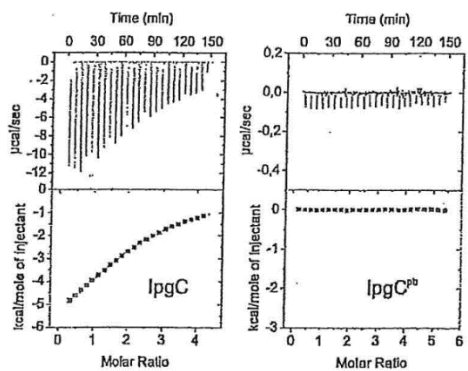

**Fig. 14.** Fig. 4D in Lunelli et al. (2009). Calorimetric titration of IpgC (invasion plasmid gene C) and IpgC$^{pb}$ (IpgC with a disrupted peptide-binding site) with IpaB peptide (PELKAP). Original data kindly provided by Michael Kolbe.

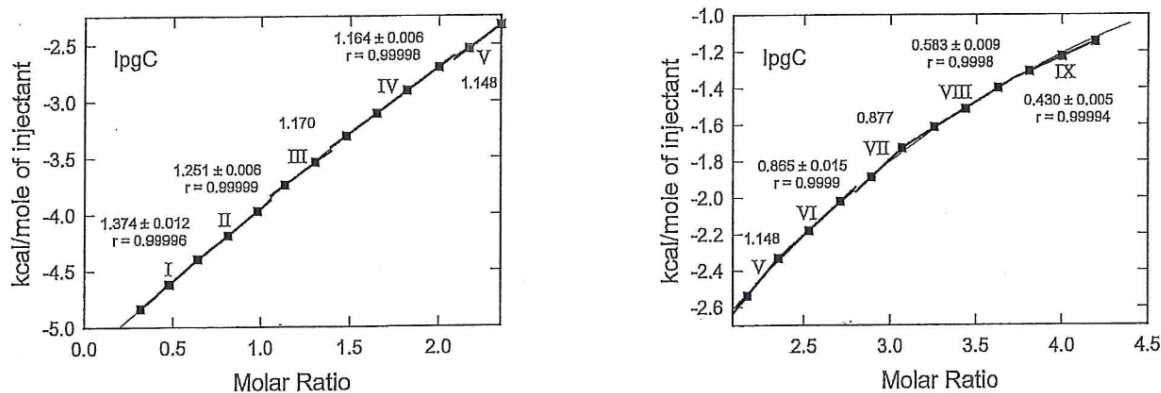

**Fig. 15.** Multiphasic profile for IpgC shown in two overlapping plots. The profile can be very well represented by 9 phases, with transitions at a molar ratio of 0.64, between 0.98 and 1.14 (jump), between 1.31 and 1.48 (jump), between 2.00 and 2.17 (jump), at 2.35, between 2.71 and 2.89 (jump), and at 3.07 and 3.72. Lines III-V are parallel, as are lines VI and VII. As shown by the exceedingly high r values for the straight lines (the lowest value, for line VIII, is 0.9998), the data are very precise.

The data are also well represented by the curvilinear profile (the thin and the thick lines are virtually indistinguishable in the above plots). It is clear, however, that the multiphasic profile does not merely represent the fitting of straight lines to sections of the curvilinear line. If so, the slopes should have decreased steadily with increasing molar ratios and there should have been no jumps. Also, the r values for lines IV , VI, VIII and IX are higher than the r values for straight lines fitted to the same sections of the curvilinear line.

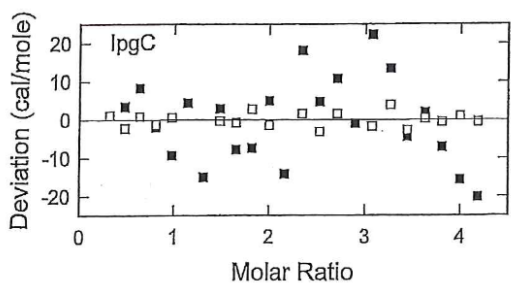

**Fig. 16**. Plot of deviates for the IpgC data.·The deviations (open squares) from the multiphasic profile are all smaller than the deviations (filled squares) from the curvilinear profile. The average absolute deviation is almost 6 times larger for the curvilinear profile (8.70 vs. 1.48 cal/mole).

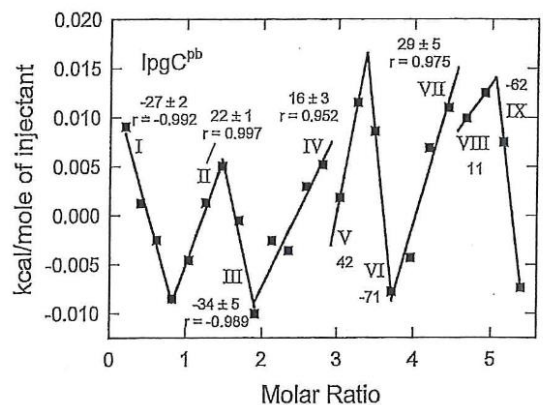

**Fig. 17.** The data for IpgC[pb] in Fig. 14 have been depicted as an essentially horizontal scatter plot. However, amplification of the y axis shows that the profile is multiphasic, with a total of 9 phases. (This is the same number of phases as in the profile for IpgC. Whether this is coincidental is not clear.) The transitions are at molar ratios of 0.83, 1.47 and 1.91, between 2.80 and 3.03 (jump), at 3.38 and 3.72, between 4.43 and 4.67 (jump), and at 5.05. The absolute r values for lines I-III are quite high, indicating that this part of the profile, and probably also the rest of the profile, has been resolved correctly

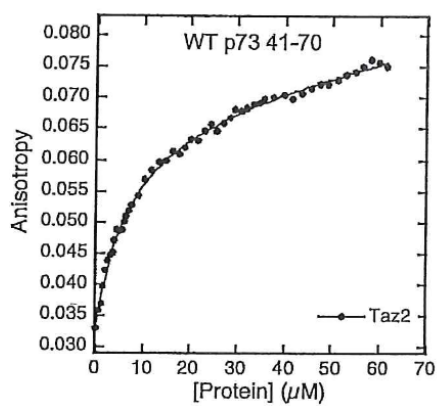

**Fig. 18.** Fig. 2 (lower right panel) in Burge et al. (2009). Fluorescence anisotropy titration of the 41-70 N-terminalpeptide of p73 to the p300 domain Taz2. Original data kindly provided by Alan R. Fersht.

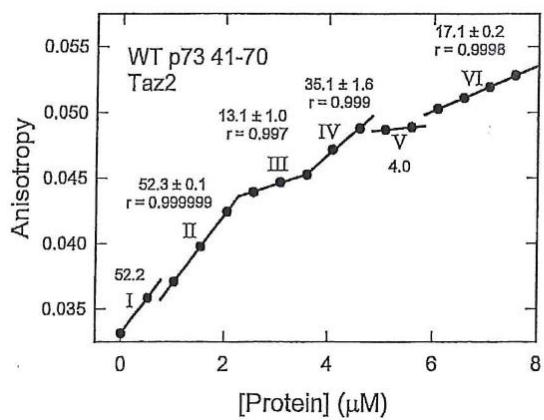

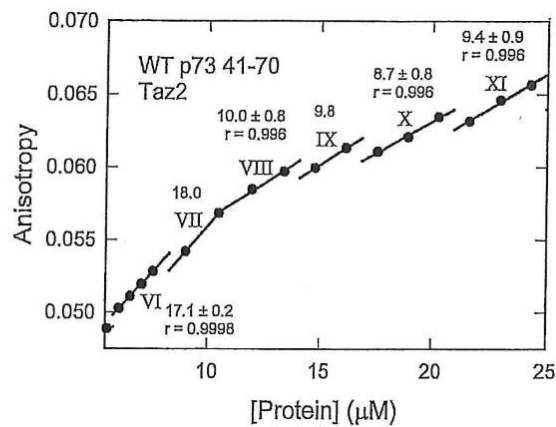

**Fig. 19** (continued on next page).

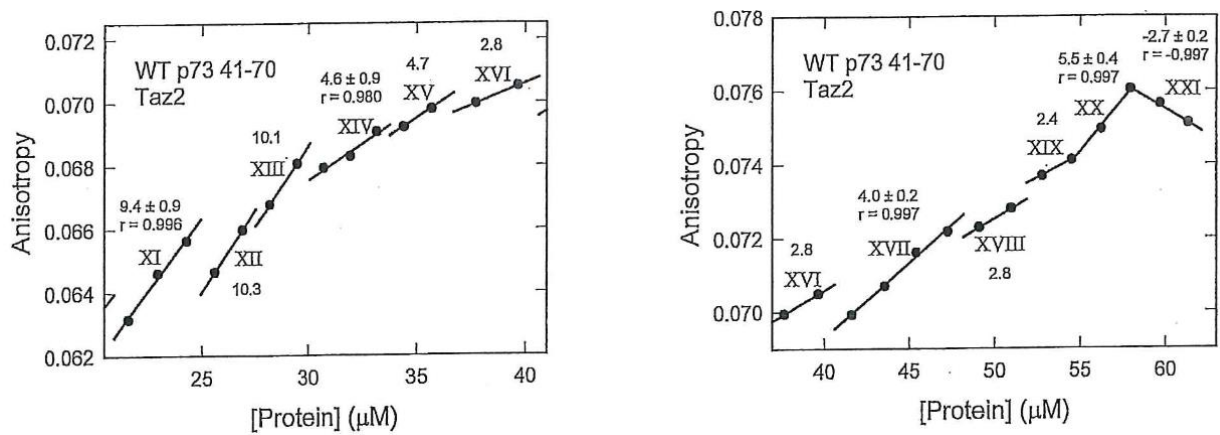

**Fig. 19** (continued from previous page). Reanalysis of the data in Fig. 18 in four overlapping plots. The data can be well represented by a total of 21 phases, with the transitions between protein concentrations of 0.5 and 1.0 μM (jump), at 2.3 and 3.6, between 4.6 and 5.1 (jump), between 5.6 and 6.1 (jump), between 7.6 and 9.0 (jump), at 10.5, between 13.4 and 14.8 (jump), between 16.2 and 17.6 (jump), between 20.3 and 21.6 (jump), between 24.3 and 25.6 (jump), between 26.9 and 28.2 (jump), between 29.5 and 30.7 (jump), between 33.2 and 34.4 (jump), between 35.7 and 37.7 (jump), between 39.6 and 41.6 (jump), between 47.3 and 49.1 (jump), between 51.0 and 52.7 (jump), and at 54.5 and 58.0. Eleven of the lines have 3 or more points. The r value for line XIV is relatively low (0.980), but the absolute r values for the other lines range from high to exceedingly high (0.996 to 0.999999). Adjacent lines are often parallel or nearly so. With a total of 21 lines based on 51 points there will necessarily be several lines (9) with only two points. However, with the exception of line V, all the 2-point lines are parallel with one or more of the other lines. Thus, it appears that the assignment of points to phases is unambiguous and that the profile has been resolved correctly. Lines I-XX have all positive slopes, whereas line XXI has a negative slope, almost significantly so (95% confidence limits: ± 2.9).

**Table 1.** Parallel lines in Fig. 19

```
                        Parallel lines:

Line    Slope           Line    Slope           Line    Slope

I       52.2            VIII    10.0 ± 0.8      XIV     4.6 ± 0.9
II      52.3 ± 0.1      IX      9.8             XV      4.7
                        X       8.7 ± 0.8
VI      17.1 ± 0.2      XI      9.4 ± 0.9       XVI     2.8
VII     18.0            XII     10.3            XVIII   2.8
                        XIII    10.1            XIX     2.4

Except for the last group, all lines are adjacent.
```

**pH**

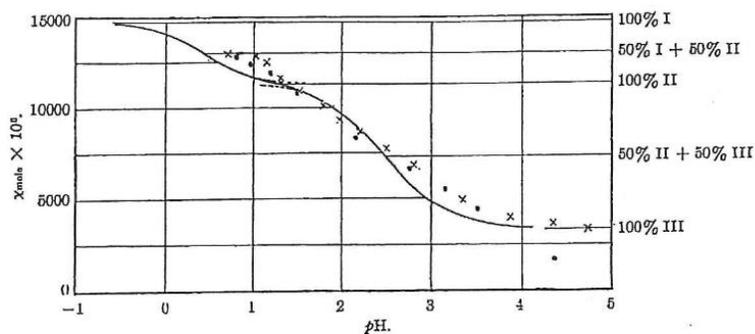

**Fig. 20.** Fig. 1 in Theorell (1941). Data from Table I (cytochrome) and Table II (ferricytochrome)

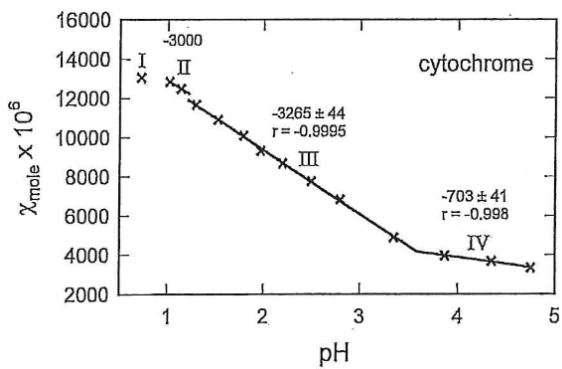

**Fig. 21.** Tetraphasic profile for data in Table I. Transitions between pH 0.72 and 1.02, between 1.14 and 1.30 (jump), and at 3.59. Lines II and III are about parallel. Very high absolute r value for the 8-point line III

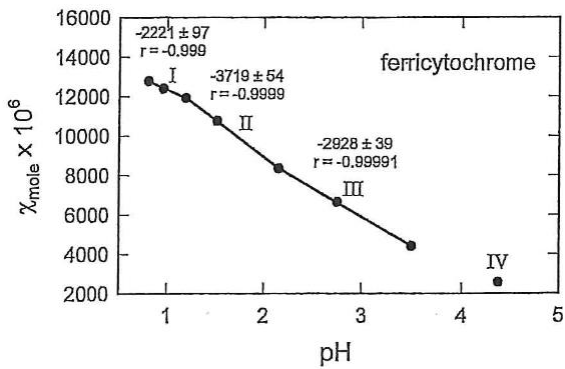

**Fig. 22.** Tetraphasic profile for data in Table II. Transitions at pH 1.19 and 2.15, and between 3.50 and 4.37. High to very high absolute r values for lines I-III.

The profiles are clearly multiphasic, not curvilinear as represented in Fig. 20.

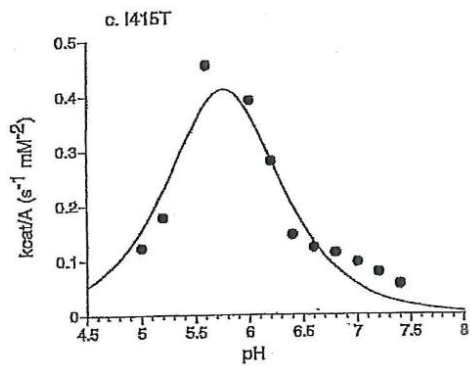

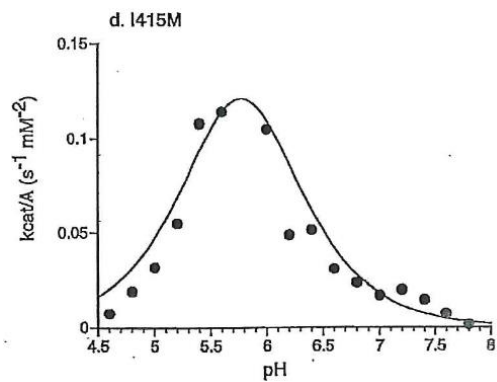

**Fig. 23** (left above). Fig. S1Bc in Guo et al. (1998). pH profile for yeast pyruvate decarboxylase. (See paper for definition of $K_{cat}/A$.)

**Fig. 24** (right above). Fig. S1Bd in Guo et al. (1998). pH profile for yeast pyruvate decarboxylase.

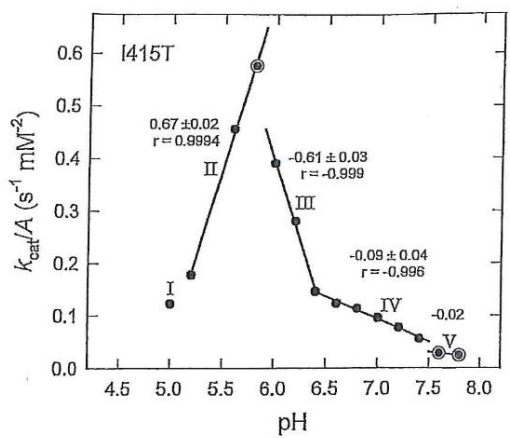

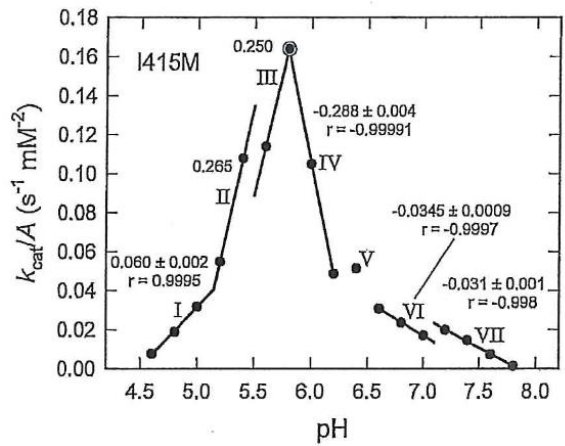

**Fig. 25** (left above). Data from Table S5 in Guo et al. (1998). The circled points at pH 5.8, 7.6 and 7.8 were omitted from Fig. 23. (At pH 5.4 a point with an y value of 1.28 was correctly omitted.) The profile can be well represented as pentaphasic, with the transitions between pH 5.0 and 5.2, between 5.8 and 6.0 (jump), at 6.4, and between 7.4 and 7.6 (jump). The absolute r values for lines II-IV are high.

**Fig. 26** (right above). Data from Table S5 in Guo et al. (1998). The circled point at pH 5.8 was omitted from Fig. 24. The profile can be well represented as heptaphasic, with the transitions at pH 5.15, between 5.4 and 5.6 (jump), at 5.8, between 6.2 and 6.4, between 6.4 and 6.6, and between 7.0 and 7.2 (jump). The data are insufficiently detailed in the range of phase V for resolution of the line. The absolute r values for lines I, IV, VI and VII are high to very high. Lines II and III are parallel, as are lines VI and VII.

The profiles are clearly multiphasic, with straight and, in part, parallel lines. The sharp transitions and discontinuities rule out the authors' bell-shaped profiles which give very poor fits even after omission of the points at pH 5.8.

# Folding/unfolding

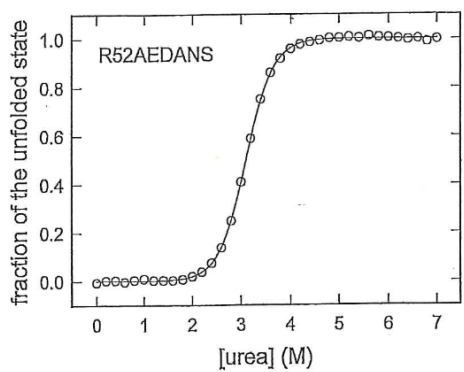

**Fig. 27.** Urea-induced equilibrium unfolding of R52AEDANS (see Arai et al. (2011) for the structure of this mutant of dihydrofolate reductase (DHFR)). For clarity, only the data and the fitting curve to a two-state transition for R52AEDANS in Fig. 2c are shown. Original data kindly provided by Munehito Arai.

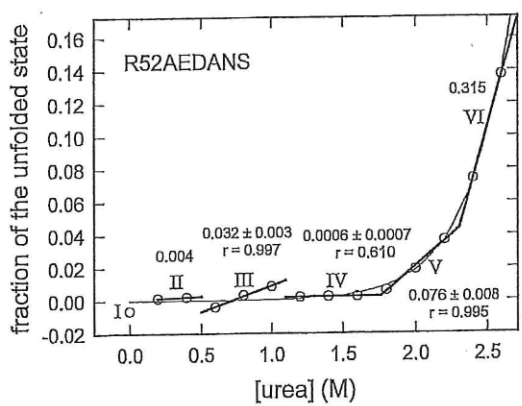

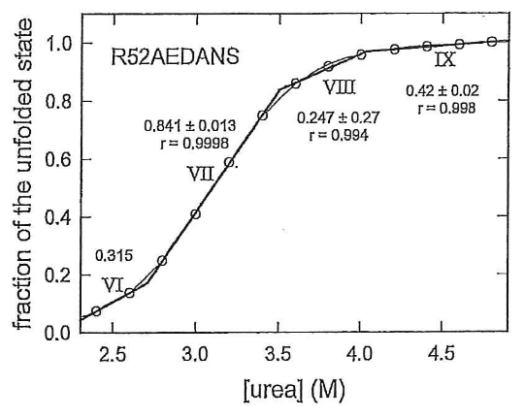

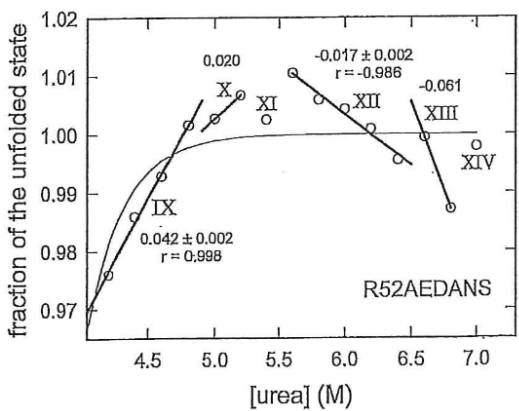

**Fig. 28.** Multiphasic profile for the data in Fig. 27 shown in three overlapping plots. The data can be well represented by 14 phases with the transitions between 0 and 0.2 M urea, between 0.4 and 0.6 (jump), between 1.0 and 1.2 (jump), at 1.76, 2.30, 2.71, 3.50, 4.03 and 4.9, between 5.2 and 5.4, between 5.4 and 5.6, between 6.4 and 6.6 (jump), and between 6.8 and 7.0. The data are insufficiently detailed in the range of phase XI for resolution of the line. The slopes are negative for lines XII and XIII. The absolute r values for lines III, V, VII, VIII, IX and XII range from quite high to very high.

The fits to the curve in Fig. 27 are good and may appear to be perfect. However, when the y axis is expanded at low and especially at high urea concentrations, it becomes clear that the multiphasic profile gives much better fits.

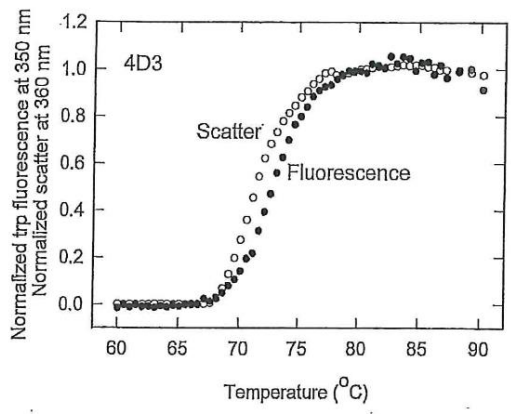

**Fig. 29.** Unfolding (intrinsic tryptophan fluorescence) and aggregation (scattering) of 4D3 lipase. Both measurements were made on the same protein sample simultaneously. Fig. S4b in Kamal et al. (2011) redrawn for clarity. Original data kindly provided by Shoeb Ahmad.

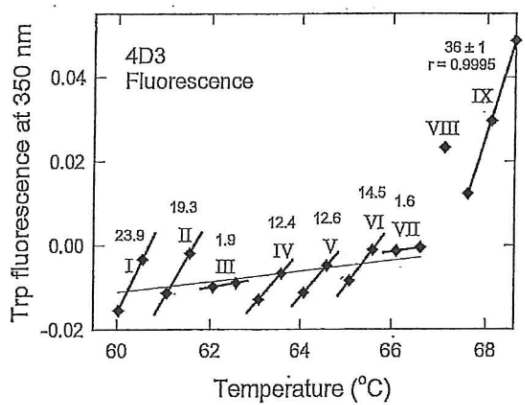

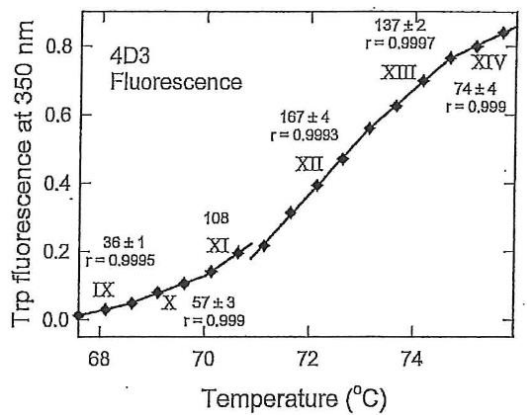

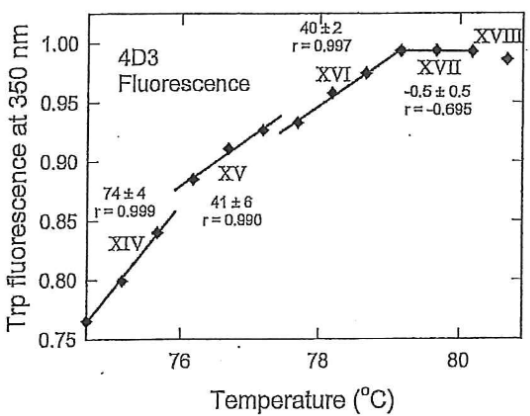

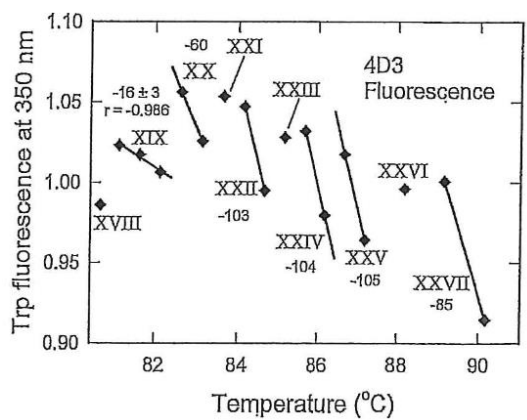

**Fig. 30.** Multiphasic profile for fluorescence from 60 to 90°C. As shown in four overlapping plots, the profile can be well represented by a total of 27 phases, with the transitions mostly in the form of jumps. The data are insufficiently detailed for phases VIII, XVIII, XXI and XXVI for the lines to be resolved. Lines III and VII are precisely parallel, as are lines IV and V, lines XV and XVI, and lines XXII, XXIV and XXV. Line XVII is horizontal, and lines XIX, XX, XXII, XXIV, XXV and XXVII have negative slopes. The first 7 phases may as indicated be fitted by a single straight line, but the deviations from this line are large, on the average 7.3-fold greater than the deviations from line IX. Lines IX, X, XII, XIII and XIV have r values of 0.999 or higher.

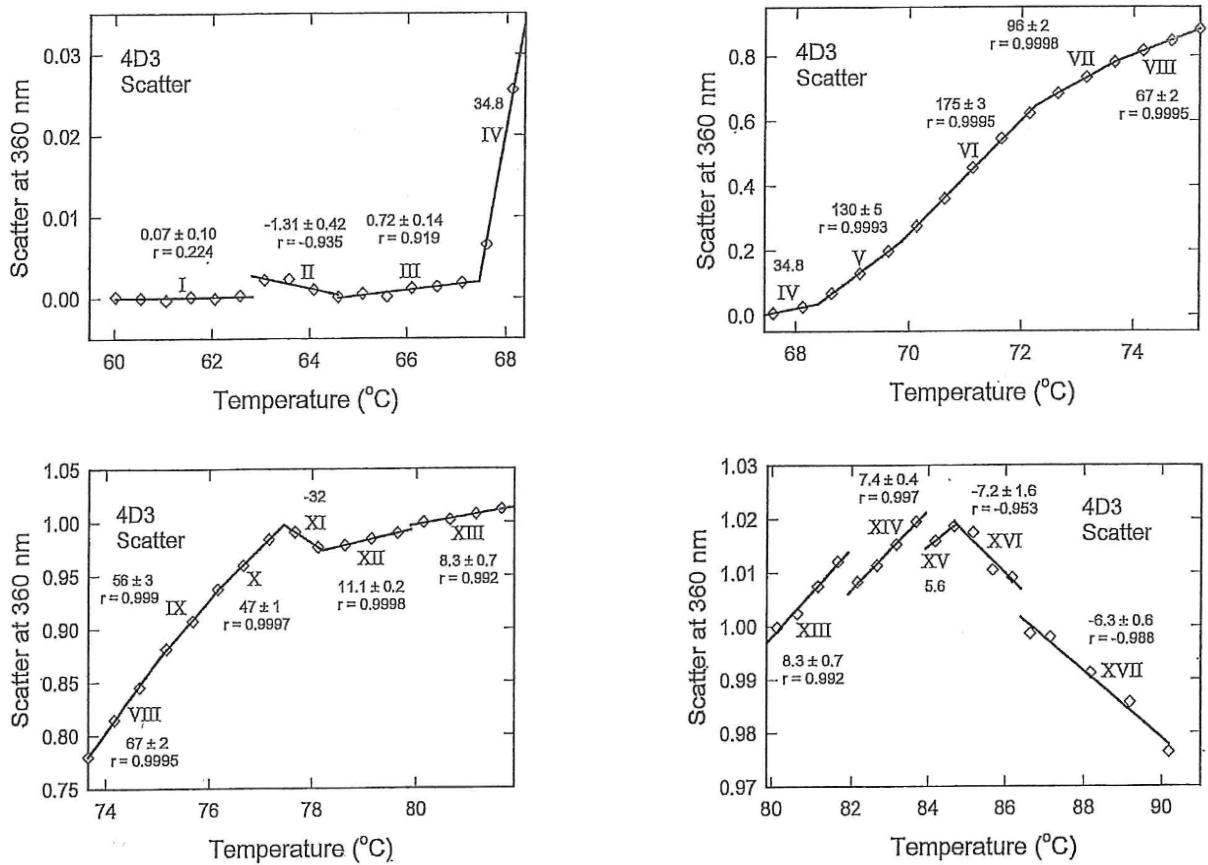

**Fig. 31.** Multiphasic profile for scatter from 60 to 90°C. As shown in four overlapping plots, the profile can be well represented by a total of 17 phases, with the transitions between 62.6 and 63.1°C (jump), at 64.6, 67.5, 68.4, 69.8, 72.3, 73.7, 75.2, 76.2, 77.5 and 78.2, between 79.6 and 80.1 (jump), between 81.6 and 82.2 (jump), between 83.7 and 84.2 (jump), at 84.7, and between 86.2 and 86.6 (jump). Lines XIII-XV are parallel or approximately so, as are lines XVI and XVII. Lines II, XI, XVI and XVII have negative slopes. Lines V-X and XII have r values of 0.999 or higher.

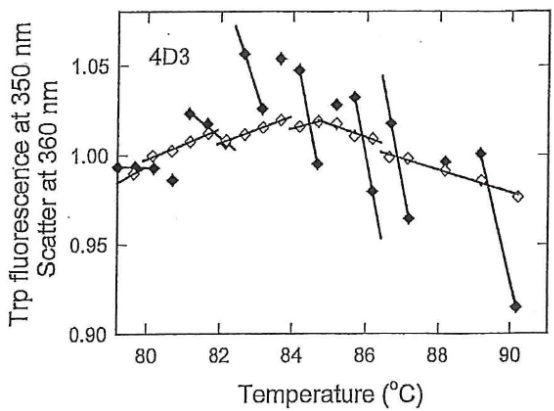

**Fig. 32.** The two measurements were made on the same protein sample at the same time, and the overall pattern appears to be the same (Fig. 29). Yet, the multiphasic profiles differ dramatically, as shown here at high temperatures.

# Chain length

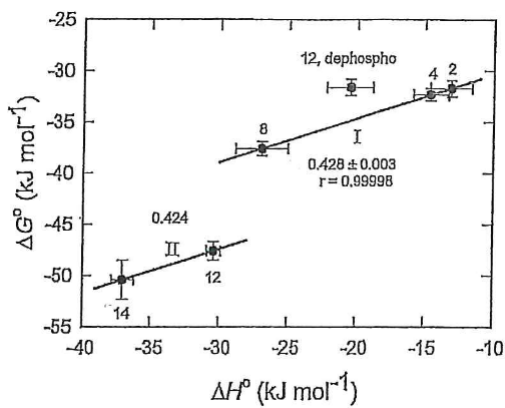

**Fig. 33.** Binding of acyl-CoA to acyl-coenzyme A binding protein. Data from Table 3 in Færgeman et al. (1996). The ligands group on two precisely parallel lines. The 12-dephospho ligand is outside these lines. The r value for line I is exceedingly high. A plot of $T\Delta S^o$ vs. $\Delta H^o$ gives the same pattern (not shown). Apparently, an abrupt conformational change of some kind has caused the marked jump between n = 8 and n = 12 as well as the discontinuous transitions in the other plots for chain length.

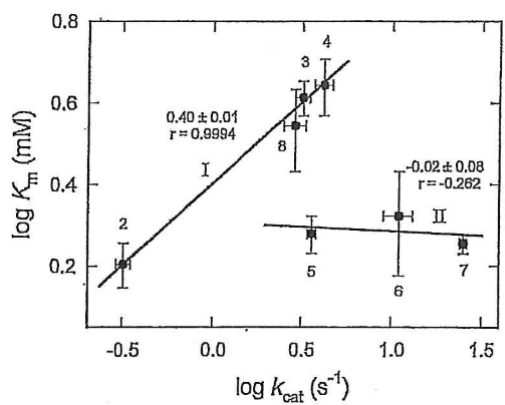

**Fig. 34.** Activity of *Yersinia* protein-tyrosine phosphatase. Data from Table II in Dunn et al. (1996). Substrates $((Glu)_4\text{-}NH\text{-}(CH_2)_n\text{-}OPO_3^{2-})$ with 2-4 methylene groups form a precisely straight line in a log-log plot of the kinetic constants. Substrates with 5-7 methylene groups can be represented by a horizontal line separated from line I by a marked jump. The transition from 7 to 8 methylene groups appears to be in the form of another marked jump. In the plot, the substrate with 8 metylene groups is taken to be off line I. (Its inclusion will lower the r value to 0.982.)

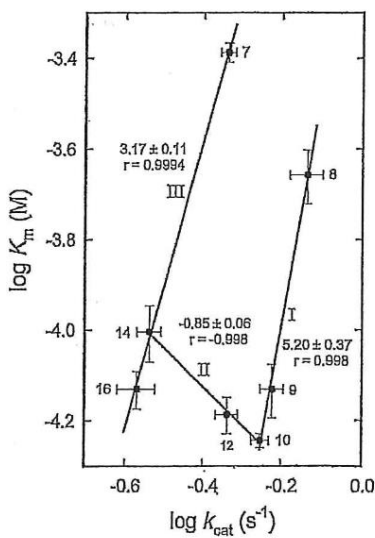

**Fig. 35.** Substrate selectivity of fatty acid amide hydrolase. Data from Table 1 in Patricelli and Cravatt (2001). Increasing the length of unbranched p-nitroanilide substrates from 8 to 10 carbons gave a straight line in a log-log plot of the kinetic constants and caused the affinity to increase and $k_{cat}$ to decrease. Increasing the length from 10 to 14 caused a further decrease in $k_{cat}$, but now the affinity decreased. In addition to the discontinuity at a chain length of 10, there are also marked discontinuities between 7 and 8 and at 14. Substrates with 7, 14 and 16 carbons form a precisely straight line, but this line could possibly be accidental.

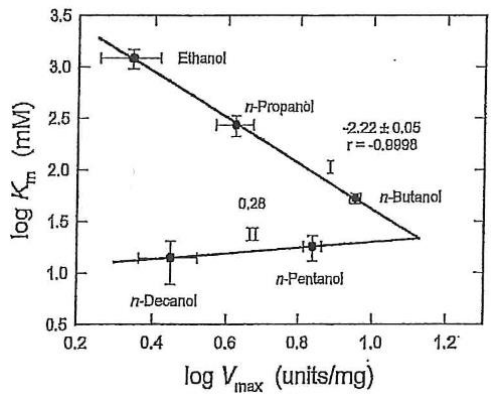

**Fig. 36.** Oxidation of alcohol substrates by ERAB/L-3-hydroxyacyl-coenzyme A dehydrogenase type II. Data from Table I in Yan et al. (1999). Increasing the chain length from 2 to 4 gave a straight line in a log-log plot of the kinetic constant, with a high absolute r value, and with increasing affinity and $V_{max}$. Increasing the chain length still further caused a sharp transition. The affinity continued to increase, but $V_{max}$ was much reduced.

## Conclusions

The profiles reanalyzed in this paper can all be well represented by multiphasic profiles. The lines are straight, often with very to exceedingly high r values (see Figs 7, 13, 15 (both plots), 19 (first plot), 22, 26 and 33 for lines with absolute r values of 0.9999 or higher). Adjacent lines are quite often parallel as seen in Figs 3, 6, 7, 15 (both plots), 19, 21, 26, 30 (plots 1, 3 and 4), 31 (plot 4) and 33. Especially remarkable is the large number of parallel lines in the anisotropy data in Fig. 18, see Fig. 19 and Table 1, and in the unfolding data in Fig. 29, see Fig. 30 (plots 1 and 4). Many non-parallel lines are also separated by marked to very large jumps. In contrast to the good fits to multiphasic profiles, the authors' curvilinear interpretations often give poor or very poor fits (Figs 2-4, 6-7, 10-13, 23-24, and 28 (plot 1 and 3). Conventional plots may sometimes appear to give good fits, as e.g. for the ITC data in Fig 14 (left), but a comparison of fits (Fig. 16) shows the deviations to be much larger than for the multiphasic profile. The anisotropy data in Fig. 18 are also much better represented as multiphasic (see Fig. 19, all four plots), as are the data for urea-induced unfolding in Fig. 27 (see Fig. 28, plot 1 and 3). In addition to the poor fits, points are often unevenly distributed around the curvilinear profile (Figs 2, 6 and 7).

In addition to the current-voltage data in Figs 1 and 5, a large number (several hundred) of similar data sets has been reanalyzed. If the data are detailed and precise, the profiles are multiphasic, not curvilinear. The same numbers and conclusion hold for data for binding as studied by ITC and for data for pH profiles. For the latter data compare the authors' Figs 23 and 24 with the multiphasic interpretation in Figs 25 and 26. Presumably in order to obtain better fits to bell-shaped curves, the authors have simply here and elsewhere omitted point(s) that give poor fits, with no mentioning of this in the text. Unfortunately, it has not been possible to get *Biochemistry* to correct these misrepresentations. The fluorescence and scattering data in Fig. 29 can both be well represented by multiphasic profiles (Figs 30 and 31), but it is remarkable that the two measurements of the same sample give widely different profiles, both with respect to the number of phases and to the slopes of the lines (Fig. 32). The profiles for the effect of chain length (Figs 33-36) are clearly bi- to tetraphasic, with parallel lines (Fig. 33) and with sharp transitions and large jumps.

The present data are all for biological systems. However, data for some non-biological systems can also be well represented by multiphasic profiles. Thus, pH profiles for different chemical reactions can be better represented by multiphasic than by curvilinear profiles (in preparation, to be submitted to arXiv). Also, unpublished data for apparently very simple systems (injection of low concentrations of methanol or ethanol (in water) into water, or injection of propanol into methanol, in the absence of water), can be well represented by multiphasic profiles, with several parallel lines.

The multiphasic profiles require explanation. Obviously, the so-called Boltzmann function or other continuous functions cannot underlie what is happening. Here I propose that the inherent noise-free process is multistate and is not described by a continuous function. Other explanations are no doubt possible, but sudden changes of state are a characteristic of many physical systems, although these tend not to be well known to biologists. The *average* curves for binding or other processes obscure the essentially discontinuous underlying process. (It is unclear why the averages often give good fits to continuous functions.) Abrupt changes of state may be involved in the sudden opening and closing of ionic channels (Kaufman et al. 2014, Lin and Eisenberg 2014). Is there one underlying cause for multiphasic profiles in the many very different systems, or can the profiles arise in different ways? It seems most unlikely that multiphasic profiles can be artificial, the systems and experimental setups are simply too hugely different. In view of the finding that the profiles are (invariably) better represented as multiphasic than as curvilinear, it seems that the use of a single continuous equation to represent what is happening over a wide range of the independent variable should be reconsidered. To be sure, there may be reasons why the fits to the conventional curves are imperfect, but these reasons do not explain why the fits to multiphasic profiles are so good. The better fits to multiphasic profiles mean that their data are more precise, usually very much so, than the authors are aware of. For correct results and to avoid misinterpretations it seems imperative that all data should be analyzed as rigorously as possible.

**Acknowledgments** – I am very grateful to Bob Eisenberg for his invaluable advice, help and encouragement without which the publication of these findings would not have been possible. I am also indebted to Lars Helge Frivold for his invaluable help in the preparation of the manuscript.